\documentclass{ws-book9x6}

\usepackage{xcolor}
\definecolor{ceccogreen}{rgb}{0.1, 0.85, 0.2}

\begin{document}

\renewcommand\thepage{\roman{page}}

\tableofcontents

\setcounter{page}{1}

\chapter{The Jamming Transition and the Marginally Stable Solid}\label{jamming}

\author[F. Arceri, E. I. Corwin, C. S. O'Hern]{Francesco Arceri$^1$, Eric I. Corwin$^2$, Corey S. O'Hern$^1, ^3, ^4$ }
\address{\textsuperscript{$1$}Department of Mechanical Engineering and Materials Science, Yale University, New Haven, Connecticut 06511, USA \\ \textsuperscript{$2$}Department of Physics and Materials Science Institute, University of Oregon, Eugene, Oregon 97403, USA \\
\textsuperscript{$3$}Department of Physics, Yale University, New Haven, Connecticut 06511, USA \\
\textsuperscript{$4$}Department of Applied Physics, Yale University, New Haven, Connecticut 06511, USA}

\section{Introduction}
Few things in this world are as common as earth, dirt, and sand.  And yet, even as the most exotic, the most distant, and the smallest objects in our universe have fallen prey to human understanding, the nature and origin of the structural and mechanical properties of these materials has been, until recently, surprisingly resistant.  It was not for a lack of trying.  The \textit{jamming} transition, by which a collection of macroscopic particles goes from flowing to rigid, lies not only at the heart of building practices throughout human history, but is ubiquitous in everyday life. Examples range from grains poured into a container~\cite{behringer_physics_2019}, to foams and emulsions in foods such as ice cream and mayonnaise~\cite{van_hecke_jamming_2010-1}, and to sand piles and gravel at the beach and in city parks~\cite{cates_jamming_1998}.  The jamming transition from fluid-to-solid behavior in these different systems shares a fundamental feature: the energy scale of the interaction between particles is sufficiently large that thermal fluctuations at room temperature are far too small to affect the dynamics. The control parameters of the jamming transition are therefore the external pressure exerted on the system and the volume in which the system is confined. The first academic study of jamming dates to 1727, when Reverend Steven Hales studied the structures formed by the contacts between dried peas, compressed in an iron pot~\cite{hales_vegetable_1727-1}.  Such luminaries as Isaac Newton~\cite{newton_newtons_1848} and James Clerk Maxwell~\cite{maxwell_calculation_1864} attempted to analytically solve for the properties of jammed packings, but were only able to determine bounds on number of contacts per particle required for mechanical stability.  More recently, Bernal connected the amorphous geometric structure of spheres to the properties of the liquid state of matter \cite{bernal_geometry_1960-3}.  However, it was not until nearly the turn of this century that the modern study of jamming was initiated~\cite{wittmer_explanation_1996,de_gennes_granular_1999-1,tkachenko_stress_1999}.  Particularly, Liu and Nagel~\cite{liu_jamming_1998} made the critical realization that athermal jamming could be united with another age old problem: the glass transition.  Only within the last decade has a first-principles description of jamming transition seemed possible, enabled by insights from the mean-field theory of glasses and jamming~\cite{parisi_theory_2020}.  Because in this framework the jamming transition takes place within a full replica symmetry breaking (RSB) phase~\cite{mezard_nature_1984}, this seemingly simple problem has revealed a world of amazing complexity.

Recent developments in the physics of the glass transition have led to groundbreaking results in the field of jamming. Although the glass transition signals a drastic (and markedly distinct) dynamical slowdown upon cooling, both transitions can be observed in systems of hard particles, which do not deform during collisions. The simple \textit{hard-sphere} model, in particular, has offered theoretical physicists fertile ground for building a mean-field theory of glasses, with jamming occurring in the limit of infinite pressure. Despite the great advances that the hard-sphere model has brought to the field, the fact that it has a discontinuous inter-particle potential represents a major obstacle for computing the mechanical properties in finite-dimensional systems. As a remedy, soft-sphere models, where particle deformations are described by shared volume between particles, have been used in the field as they allow more direct calculations of inter-particle forces. The fact that the jamming transition for frictionless spherical particles can be studied using both the hard- and soft-particle models reflects its geometrical nature. By considering the average number of contacts per particle, $z$, one finds that the jamming of hard and soft spheres occurs at $z_c=2d$~\cite{maxwell_calculation_1864}, i.e., the minimum number of contacts to ensure rigidity~\cite{ohern_random_2002} (Maxwell's criterion). For soft spheres, which can be compressed beyond the jamming point, the excess number of contacts scales as:
\begin{equation}
\Delta z \equiv z - z_c \sim \Delta\varphi^{1/2},
\end{equation}
where $\varphi$ is the packing fraction and the excess packing fraction $\Delta\varphi = \varphi - \varphi_J$ represents the amount of compression above the jamming threshold, which is itself protocol dependent~\cite{durian_foam_1995,ohern_jamming_2003}. Further, several studies have shown an exact correspondence in the inter-particle separations between jammed hard-particle and soft-particle packings~\cite{ohern_jamming_2003,donev_pair_2005}, confirming that accessible configurations of hard and soft spheres are identical near jamming onset~\cite{brito_rigidity_2006,wu_response_2017,arceri_vibrational_2020-1}.

\section{Jamming Criticality}
Many numerical studies have documented critical behaviors of bulk quantities near the jamming transition, including observation of power-law scaling~\cite{ohern_random_2002,ohern_jamming_2003,ellenbroek_jammed_2009}, scaling collapse of the elastic moduli and excess contact number~\cite{ellenbroek_critical_2006,dagois-bohy_soft-sphere_2012,goodrich_jamming_2014-1,van_deen_contact_2014}, identification of diverging length scales~\cite{silbert_vibrations_2005-2,wyart_geometric_2005,goodrich_stability_2013-1,lerner_breakdown_2014-1}, and analyses of finite-size scaling~\cite{goodrich_jamming_2014-1}. Recent studies have unified these scaling relations in a single theory of jamming using a \textit{scaling ansatz}~\cite{goodrich_scaling_2016}. This approach borrows ideas from critical phenomena, such as spontaneous magnetization and density-charge waves~\cite{widom_equation_1965,middleton_critical_1993,pazmandi_revisiting_1997}, to describe scaling relations for the energy $E$, pressure $p$, excess packing fraction $\Delta \varphi$, shear stress $s$, shear strain $\epsilon$, bulk modulus $B$, shear modulus $G$, and number of particles $N$ obtained from a single state function for the elastic energy
\begin{equation}
E(\Delta Z, \Delta \varphi, \epsilon, N) = \Delta Z ^\zeta \mathcal{E}_0\left( \frac{\Delta \varphi}{\Delta Z ^{\beta_{\varphi}}},  \frac{\epsilon}{\Delta Z ^{\beta_\epsilon}}, N \Delta Z^\psi\right)
\end{equation}
and its derivatives 
\begin{equation}
p \equiv \varphi \frac{dE}{d\Delta \varphi} \; \mbox{,} \;\;\;  s \equiv \epsilon \frac{dE}{d \epsilon} \; \mbox{,} \;\;\; B \equiv \frac{\varphi^2}{2}\frac{d^2E}{d\Delta \varphi^2} \; \mbox{,} \;\;\;  G \equiv \frac{d^2E}{d\epsilon^2} \; \mbox{.}
\end{equation}
Notice that $\Delta Z = N \Delta z$ is the total number of excess contacts in a jammed soft-sphere system. These relations yield a set of equations that couple the critical exponents. The picture is completed by the addition of the \textit{pressure-shear stress exponent equality}, which dictates that $s^2$ vanishes in the infinite-size limit as $1/N$.

Although the scaling ansatz offers a description in the $\Delta \varphi \epsilon N$ ensemble, numerical studies are often conducted in the $p \epsilon N$ or $p s N$ ensembles as both pressure and shear stress vanish at the jamming point. The scaling ansatz can be extended to the $p s N$ ensemble at finite temperature by defining a new state function for the free energy: 
\begin{equation}
 F(\Delta Z, p, s, N, T) = \Delta Z ^\zeta \mathcal{F}_0\left( \frac{p}{\Delta Z ^{\delta_{p}}},  \frac{s}{\Delta Z ^{\delta_s}}, N \Delta Z^\psi, \frac{T}{\Delta Z^{\delta_T}}\right).
\end{equation}
The new scaling exponent equation, $\delta_T = \zeta = 4$, is consistent with the scaling of the critical temperature, $T^\ast \simeq \Delta Z^4$, which defines the separation between glass and jamming-like behavior~\cite{ikeda_dynamic_2013,wu_response_2017}. A scaling theory for thermal systems is nevertheless far from complete due to glassy phenomena, such as aging and dynamical heterogeneity, which stem from the temperature-activated breaking and reformation of particle contacts, that is difficult to capture with simple scaling laws~\cite{brito_geometric_2009,schreck_repulsive_2011}. 
 
The scaling ansatz provides a framework that connects the structural and mechanical properties ($\Delta Z$ and $G$) for particle systems above the onset of jamming. The existence of such a framework implies that the jamming transition exhibits scaling invariance, a helpful tool for a renormalization group description. However, recent studies of the shear modulus near the jamming transition have highlighted the limits of the scaling ansatz~\cite{vanderwerf_pressure_2020,wang_shear_2021}. In particular, the scaling ansatz cannot explain why the shear modulus $G$ scales linearly with $\Delta Z$, a result which has been numerically tested and theoretically confirmed by effective medium theory~\cite{wyart_scaling_2010,degiuli_theory_2015}. Interestingly, the inter-particle force law does not play a more important role in determining the shear modulus in jammed packings at non-zero pressure. In addition, the scaling analysis cannot quantify deviations between the ensemble average and the large-system limit, and it does not describe local fluctuations in the elastic moduli, or that the local elastic moduli can become negative~\cite{dagois-bohy_soft-sphere_2012,wang_shear_2021}.  The importance of fluctuations in the elastic moduli suggest that we need to develop a deeper understanding of the energy landscape of jammed packings.

\section{Marginal Stability}\label{sec:theory}
\begin{figure}
    \centering
    \includegraphics[width=\columnwidth]{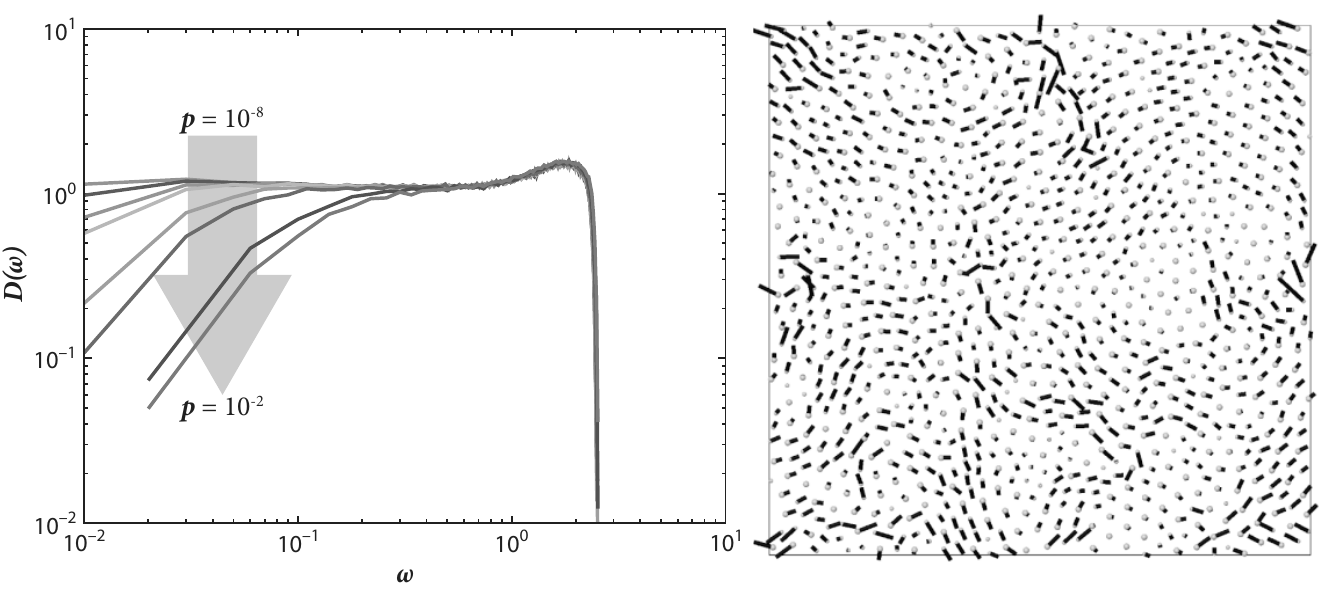}
    \caption{(a) Vibrational density of states $D(\omega)$ plotted as a function of pressure $p$ and (b) the real-space representation of a soft, extended mode. Figures reproduced with permission from Ref.~\cite{silbert_vibrations_2005-2} \copyright (2005) American Physical Society.
    }
    \label{fig:vdos}
\end{figure}

Jammed packings are \textit{isostatic} and marginally stable. In other words, they possess the minimum number of contacts to ensure mechanical stability and single bond-breaking perturbations can destabilize the entire system~\cite{wyart_marginal_2012}. Numerical and experimental studies have characterized marginal stability by measuring the vibrational density of states (VDOS) of jammed solids at finite pressure~\cite{liu_jamming_2010-1} (Fig.~\ref{fig:vdos}). In particular, the low-frequency region of the VDOS possesses \textit{soft modes}, i.e., low-frequency normal modes of vibration that are spatially extended and involve a large fraction of particles in the system. Upon decreasing the pressure of jammed soft spheres toward the unjamming transition, the number of force-bearing contacts decreases until it reaches the isostatic value, $N_{iso} = Nz/2$, where the number of contacts equals the number of constraints. Here, the low-frequency region of the VDOS develops a plateau and the frequency of the lowest mode scales as $\omega_c \sim \Delta \varphi^{1/2}$. Finite-size scaling shows that $\omega_c \rightarrow 0$ in the large-system limit, in which case, the soft modes correspond to zero-energy modes. 

A real-space interpretation of marginal stability is offered by the so-called \textit{cutting argument} introduced by Wyart, {\it et al.}~\cite{wyart_geometric_2005}. Imagine removing contacts on the edge between a subsystem of linear size $l$ and the rest of the system. If at this point the system is slightly compressed, the lack of contacts leads to a competition between the overall excess contacts $\Delta Z$ created by the compression and the missing contacts at the boundary. If the total number of contacts is equal to the isostatic value $N_{iso}$, the system possess soft modes. The number of soft modes $N_{soft}$ corresponds to the difference between the number of contacts at the boundary, which are proportional to $l^{d-1}$, and the number of extra contacts created by the compression, which scales as $\Delta Z l^d$. Therefore, a critical length $l^\ast \sim \Delta \varphi^{-1/2}$ exists, below which the system is isostatic and possesses soft modes. 
\begin{figure}
    \centering
    \includegraphics[width=0.95\columnwidth]{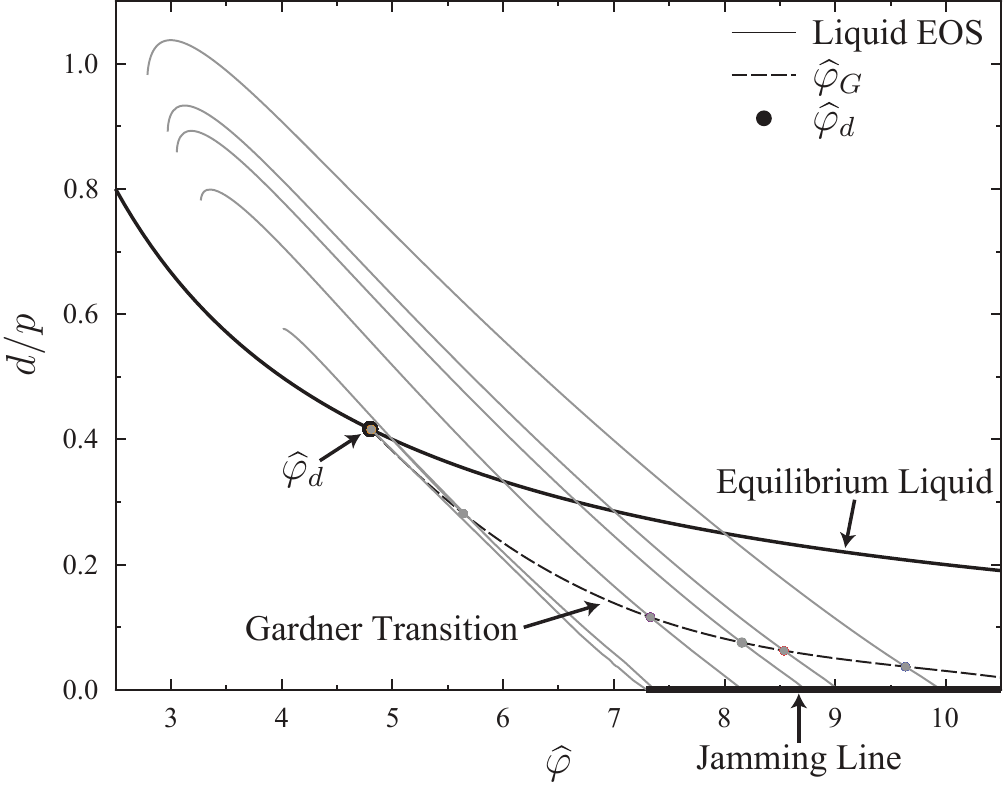}
    \caption{Glass phase diagram in the $d/p$ versus $\hat{\varphi}$ plane, where $\hat{\varphi} = 2^d \varphi /d$, $\varphi$ is the packing fraction, and $d$ is the spatial dimension. In contrast to the \textit{equilibrium liquid} equation of state (thick solid line), each thin solid line represents a state-following compression that starts from a liquid state and evolves toward the glass and RSB glass (dotted line) regions. The end-point of each state-following compression occurs on the jamming line. Figure reproduced with permission from Ref.~\cite{rainone_following_2016} \copyright (2016) IOP Publishing.}
    \label{fig:phaseDiagram}
\end{figure}

The mean-field theory of glasses provides a theoretical understanding of marginal stability in jammed solids as the jamming transition is viewed as the end-line of the glass phase diagram~\cite{parisi_theory_2020} pictured in Fig.~\ref{fig:phaseDiagram}. Here, compression of an equilibrium liquid composed of hard spheres gives rise to dynamical arrest when the system becomes confined to one of the basins of the complex and hierarchically organized free-energy landscape. Adopting the \textit{state-following formalism} developed by Rainone {\it et al.}~\cite{rainone_following_2016}, each glass state undergoes a Gardner transition upon further compression that brings the system to a marginal glass state~\cite{berthier_gardner_2019}. Here, the free-energy basins are fragmented into multiple sub-basins that are in turn fractured into sub-sub-basins \textit{ad infinitum}. These marginal glass states are separated by free-energy barriers that grow with system size and link glass formation with marginally stable packings typical of jammed materials. The mean-field theory of glasses also predicts the existence of a \textit{jamming line}, i.e., marginal glass states that cannot be further compressed occur over a range of packing fractions, which depends on the protocol used to produce the initial glass state. 

Another connection between jamming and the mean-field theory of glasses is the unjamming transition of soft spheres. Finding configurations of $d$-dimensional spheres with no inter-particle overlaps is a type of \textit{satisfiability problem}~\cite{biere_handbook_2009,mezard_information_2009,franz_universality_2017-1}. Determining whether collections of spherical particles overlap each other can be cast as a constraint satisfaction problem, where $N$ variables can be adjusted to satisfy $M$ constraints. Several algorithms have been used to solve constraint satisfaction problems, such as gradient descent~\cite{aluffi-pentini_global_1988,chacko_slow_2019}, simulated annealing~\cite{kirkpatrick_optimization_1984}, and the perceptron model~\cite{franz_simplest_2016,altieri_higher-order_2018-1}. The perceptron model has been successful in describing the high-dimensional energy landscape near the jamming transition~\cite{franz_jamming_2019}. In short, it describes a tracer particle on an $N$-dimensional hypersphere of radius $\sqrt{N}$ with $M$ obstacles in random positions placed on the hypersphere surface. The solution for the accessible configurations of the tracer particle maps onto the satisfiability problem of a jammed packing of soft spheres~\cite{franz_simplest_2016}. This theory predicts the same critical behavior of the interparticle gap and force distributions and the same scaling of the coordination number versus the pressure, $\Delta z \sim p^{1/2}$, as those found for jammed packings of spherical particles.

\section{Numerical Confirmations of the Mean-Field Theory of Jamming}
Jammed systems offer a uniquely useful arena to test the predictions of the mean-field theory.  Because they are athermal, one can devise experimental and numerical systems to directly measure the particle-scale properties that are predicted by the theory.  Over the past decade, fruitful collaborations have emerged between theory, experiment, and numerical simulation to confirm many of the mean-field theoretical predictions in athermal jammed systems. 
\begin{figure}
    \centering
    \includegraphics[width=\columnwidth]{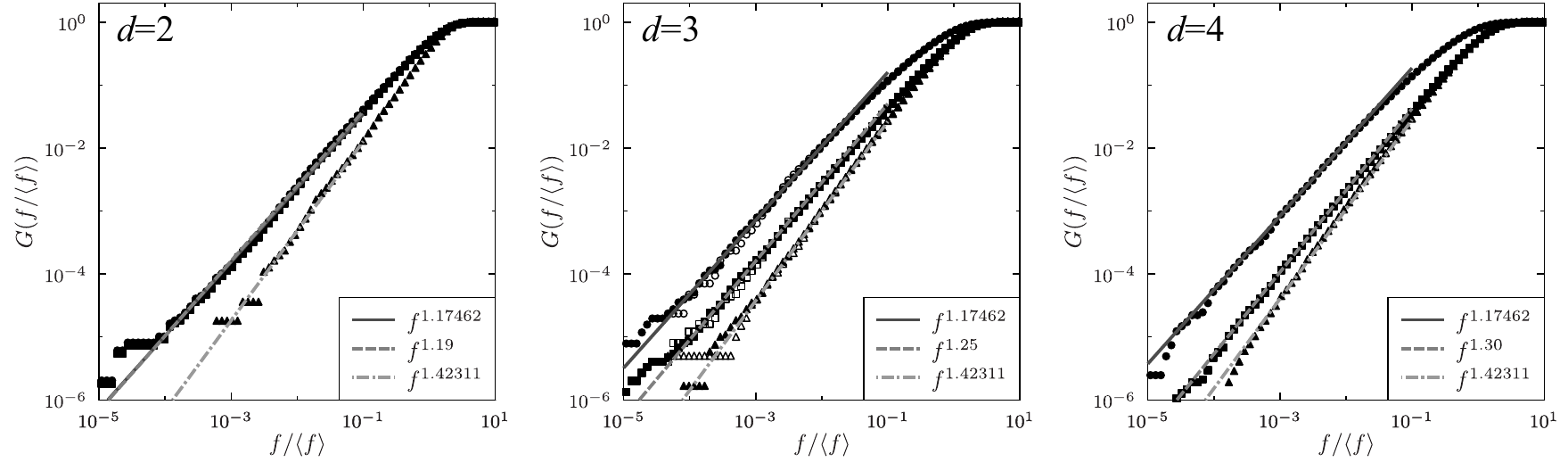}
    \caption{Cumulative force distributions $G(f)$ for $d$= 2, 3, and 4, showing power-law distributions at small forces. The overall distribution (with fits shown by the dashed lines) shows evolution with spatial dimension.  However, when split into \emph{buckler} and \emph{non-buckler} contacts, $G(f)$ for the former is well-fit by a power-law with exponent $1+ \theta_l$ and $G(f)$ for the latter is well-fit by a power-law with exponent $1+\theta_e$ in all dimensions, in excellent agreement with the mean-field theory predictions. Figure reproduced with permission from Ref.~\cite{charbonneau_jamming_2015-1} \copyright (2015) American Physical Society.}
    \label{fig:Bucklers}
\end{figure}

One of the first, and most striking, confirmations of the mean-field theory has come from the examination of the distribution of inter-particle forces, $P(f)$, in systems at jamming onset~\cite{charbonneau_jamming_2015-1}.  The small-force tail of the force distribution contributes significantly to the mechanical properties of a packing~\cite{wyart_scaling_2010}.  Both the distribution of small gaps and small forces between particles determine the statistics of contact breaking and formation when a system is mechanically perturbed.  The mean-field theory makes a precise prediction about the distribution of small forces.  If breaking a weak contact results in a spatially \textit{extended} soft mode, the tail of the force distribution scales as $P(f) \sim f^{\theta_e}$, where $\theta_e \approx 0.42311 $.  However, if breaking a weak contact results in a spatially \textit{localized} soft mode, a different argument, based on an analysis of marginal mechanical stability, predicts that the tail of the distribution will be distributed as $P(f) \sim f^{\theta_l}$, where $\theta_l \approx 0.17462$~\cite{lerner_low-energy_2013-1}.  On the face of it, the decomposition of forces into those associated with \textit{extended} and \textit{localized} excitations is a substantial task.  However, it suffices to recognize that the vast majority of localized forces can be associated with \textit{bucklers}, particles that are minimally stable (with only $d+1$ contacts), $d$ of them nearly co-planar, and hence they have only one small contact force.  As shown in Fig.~\ref{fig:Bucklers} (which presents the cumulative distribution of forces, $G(f) = \int_0^f df' P(f')$), when the forces are decomposed according to this rule, the mean-field theory predictions for the small force tails are observed in spatial dimensions all the way down to $d=2$.  This surprisingly precise agreement provided strong evidence that the mean-field results are predictive in physically relevant systems.

\begin{figure}
    \centering
    \includegraphics[width=\columnwidth]{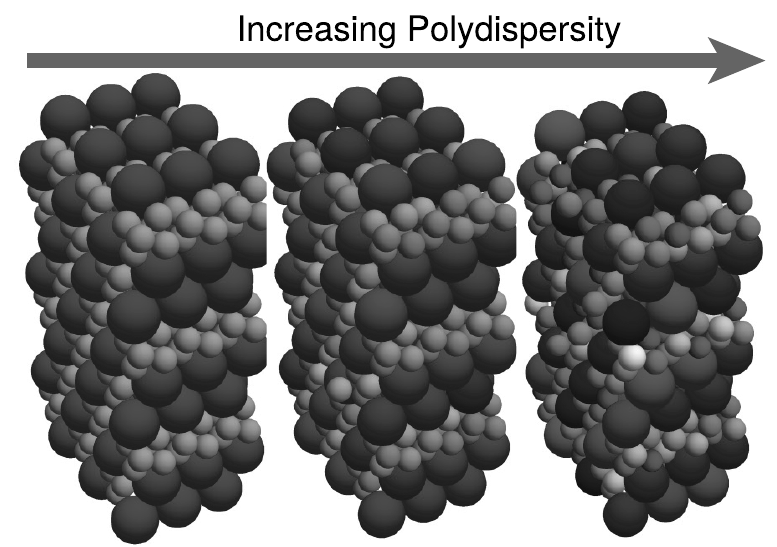}
    \caption{A crystalline sphere packing with increasing amounts of polydispersity in the particle sizes, from perfectly crystalline on the left to 5\% polydispersity on the right. Figure reproduced with permission from Ref.~ \cite{charbonneau_glassy_2019-1} \copyright (2019) American Physical Society.  }
    \label{fig:GardnerCrystal}
\end{figure}

The strength and validity of the mean-field predictions were further bolstered by a comprehensive study of the finite-size effects on jammed packings~\cite{charbonneau_finite-size_2021}.  This work showed that the aforementioned force distribution exponents are remarkably precise in low dimensions for systems as small as $N=256$ particles, showing essentially no finite-size effects.

One unexpectedly fruitful system of interest for studying the implications of the mean-field theory is the so-called \textit{Gardner crystal}, consisting of slightly polydisperse spheres packed into a nearly perfect crystal, as shown in Fig.~\ref{fig:GardnerCrystal}.  At high temperature or low pressure, the polydispersity is effectively masked by the random motions of the particles, resulting in a solid with conventional properties.  However, as the temperature is decreased or the pressure is increased, the cages around each particle shrink.  Once the gaps between particles become comparable to the scale of the polydispersity, the system is forced to make choices between a hierarchy of different possible configurations, akin to what takes place in an amorphous system \cite{charbonneau_glassy_2019-1}.
\begin{figure}
    \centering
    \includegraphics[width=\columnwidth]{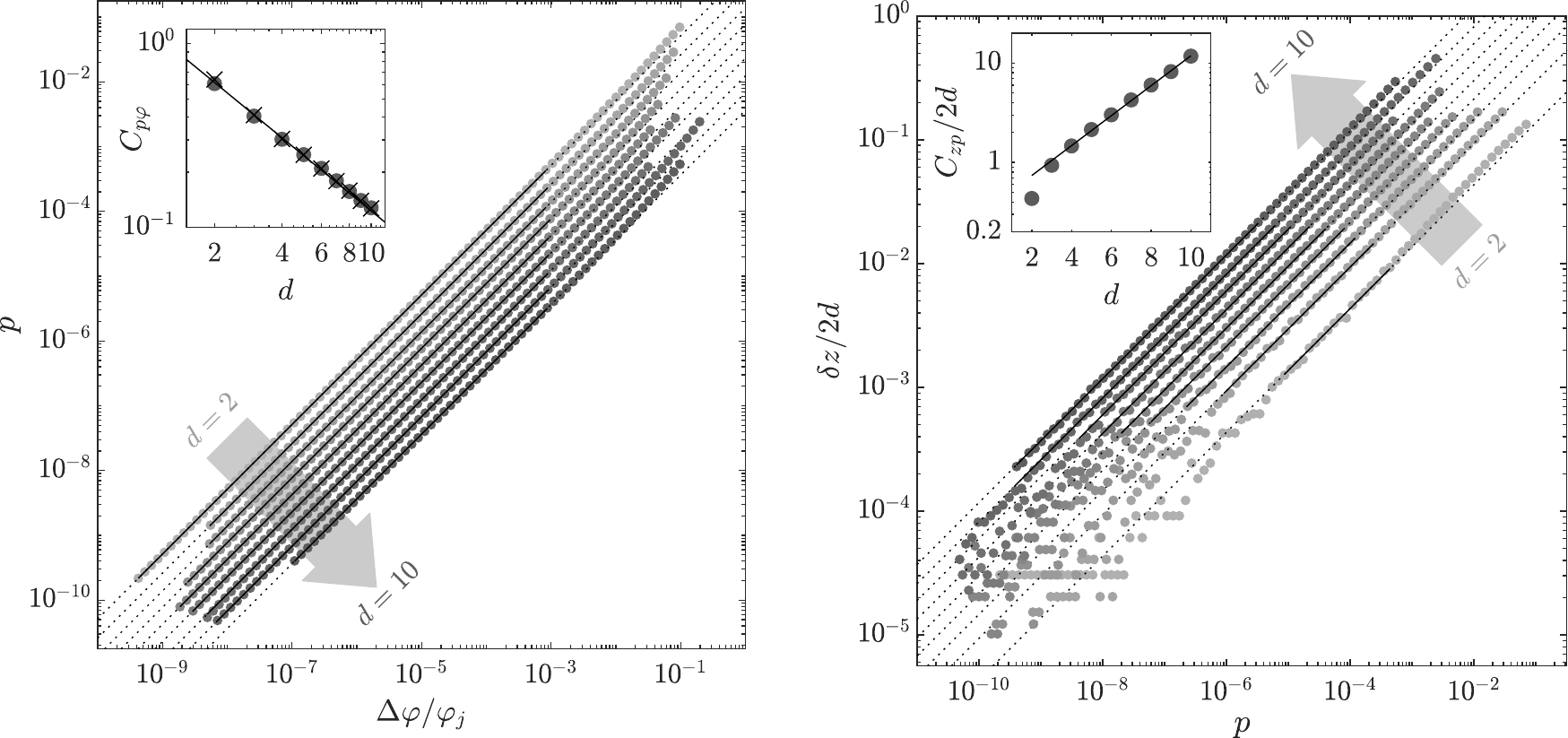}
    \caption{Scaling relations between the pressure, $p$, and the normalized excess packing fraction, $\Delta \varphi/\varphi_j$, where $\varphi_j$ is the packing fraction at jamming onset, and the excess contact number, $\delta z \equiv \Delta Z / N$ versus $p$.  The dotted lines are best fits to the scaling relations obtained by the mean-field theory of jamming, as a function of the spatial dimension from $d=2$ to $10$.  The insets show the prefactors of the scaling relations as points and the mean-field theory predictions for the prefactors as solid lines.  Figure reproduced with permission from Ref.~\cite{sartor_mean-field_2021-1} \copyright (2021) American Physical Society.}
    \label{fig:prefactors}
\end{figure}

Another direct confirmation of the mean-field theory of jamming was obtained in a study of the scaling prefactors relating the pressure, packing fraction, and number of contacts in jammed systems.  The mean-field theory provides predictions not only for the scaling exponents relating these quantities, but also for the prefactors themselves, as a function of the spatial dimension of the system~\cite{parisi_theory_2020}.  Sartor, Ridout, and Corwin demonstrated through numerical simulations that these prefactor relations hold all the way down to $d=2$ and 3~\cite{sartor_mean-field_2021-1}, as shown in Fig.~\ref{fig:prefactors}.

While the above work demonstrates the predictive powers of the mean-field theory, recent work~\cite{dennis_jamming_2020} has also directly confirmed that the energy landscape of jammed systems is consistent with full replica symmetry breaking.  As a glassy system undergoes the Gardner transition entering a marginal glass state, the energy landscape results in an ``ultrametric" structure for the very large number of marginally stable minima~\cite{gardner_spin_1985-1}. Through an exhaustive search of the nearby minima in a local region of the energy landscape of jammed packings of finite size, Dennis and Corwin were able to directly measure the degree of ultrametricity of the landscape. They found that, in the large-system limit, this landscape became precisely ultrametric, with the distance to ultrametricity scaling as $N^{-1/2}$.  Thus, jammed systems can be viewed as being located deep within the Gardner phase, as described by the mean-field theory.

\section{Experimental Validations of the Mean-Field Theory of Jamming}

\begin{figure}
    \centering
    \includegraphics[width=\columnwidth]{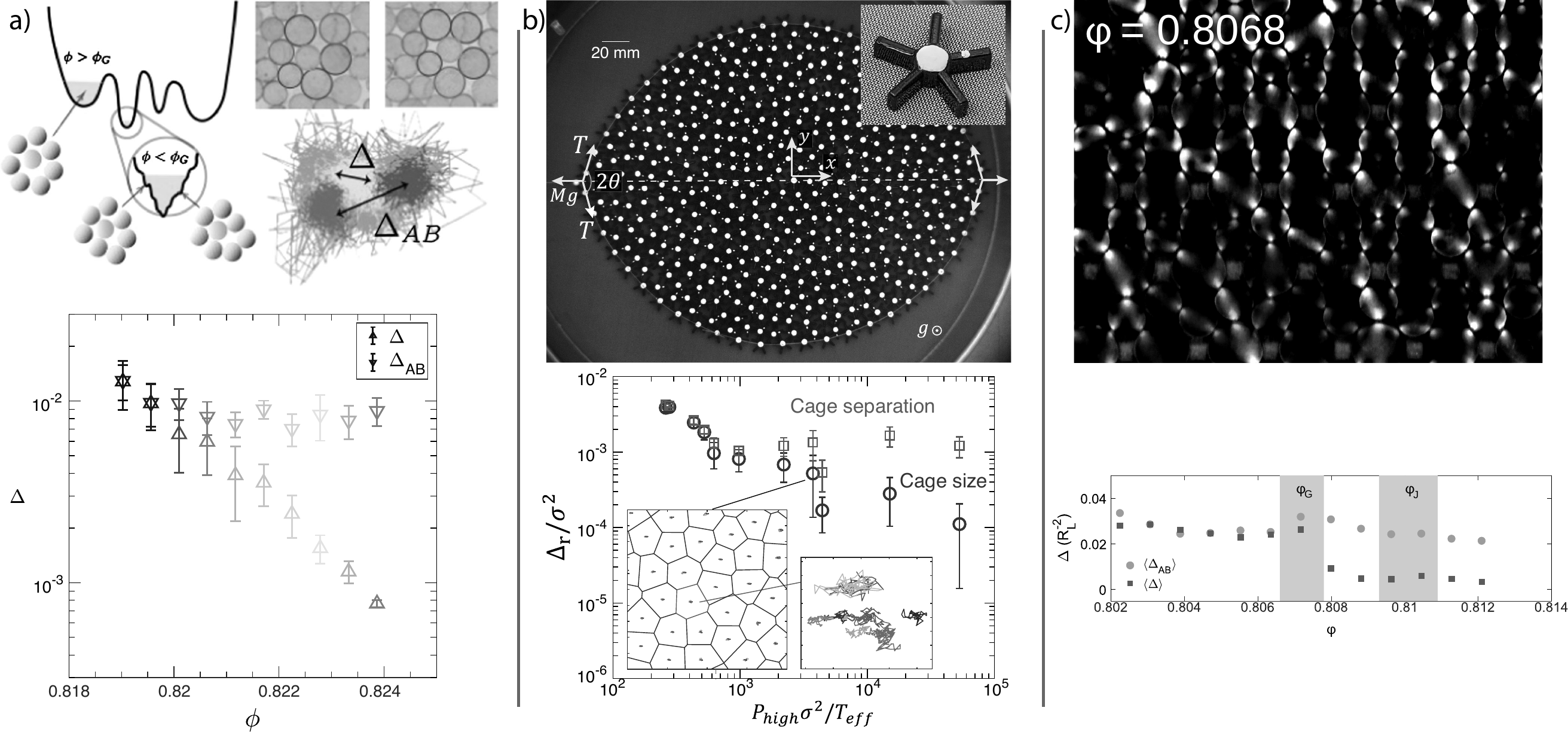}
    \caption{Experimental evidence for the Gardner transition in driven, athermal disk packings, showing the characteristic splitting between $\Delta$ and $\Delta_{AB}$ at the Gardner packing fraction or pressure.  a) Vibrated disk packings. b) Air-fluidized rotors. c) Slightly polydisperse crystalline configuration of photoelastic disks. Figures reproduced with permission from (a) Ref.~\cite{seguin_experimental_2016-2} \copyright (2016) American Physical Society, (b) Ref.~\cite{xiao_probing_2022} \copyright (2022) American Physical Society, and (c) Ref.~\cite{kool_gardner-like_2022-1} \copyright (2022) the authors.}
    \label{fig:gardnerExperiment}
\end{figure}

Experimental tests of the mean-field theory have chiefly focused on the detection of signatures of the Gardner transition in driven, athermal systems that mimic thermal systems. Seguin and Dauchot \cite{seguin_experimental_2016-2} constructed a granular system of vibrated disks and used it to explore the fracturing of the energy landscape as pressure is increased.  An initial, \textit{high-energy} amorphous configuration was created by confining a system of plastic disks in a fixed volume, and thus at a fixed packing fraction.  Energy introduced through vibration allowed the system to make and break inter-particle contacts and explore the local energy landscape.  The system was then cyclically quenched to a higher packing fraction (and thus a higher pressure) and then decompressed to the original packing fraction.  The mean squared displacement of particles, $\Delta$ at the high packing fraction, as well as the mean-squared displacement between cycles, $\Delta_{AB}$, were measured.  Figure \ref{fig:gardnerExperiment} (a) shows that these two measurements depart from one another as the quench packing fraction is increased, which is the signature of the Gardner transition in this amorphous 2D system.  Xiao, Liu, and Durian made a similar measurement on a dissimilar amorphous system, also observing the clear signature of the Gardner transition~\cite{xiao_probing_2022}.  Rather than using vibrated disks, this work instead used a 2D system of air-fluidized rotors, shown in Fig.~\ref{fig:gardnerExperiment} (b), constructed with five-fold symmetry to frustrate crystallization.  A flexible boundary made of a chain of particles subject to a fixed tension serves to precisely control the pressure of the system, allowing measurements that can be directly compared to the Gardner transition results predicted for thermal systems (see Fig.~ \ref{fig:gardnerExperiment} (b)).

The aforementioned Gardner crystal provides opportunities for experimental studies of the Gardner transition as well.  Kool, Charbonneau, and Daniels constructed a 2D system of photo-elastic disks and were able to track the formation of persistent contacts as the system passed through a Gardner-like transition \cite{kool_gardner-like_2022-1}.  This work also showed the characteristic splitting between fluctuations within a system and fluctuations between systems at the Gardner transition (see Fig.~\ref{fig:gardnerExperiment} (c)).

\section{Beyond Sphere Packings}
We have presented numerical and experimental confirmations of several predictions from the mean-field theory of jamming in the context of packings of identical frictionless spherical particles, interacting via excluded volume repulsion.  However, most of the materials that show jamming transitions are not composed of frictionless, spherical particles--think jelly beans, grains, and rocks.  Clogging transitions, which are similar to jamming transitions, also occur in crowds of people and organisms~\cite{bunde_science_2002}.  Moreover, many biological systems are extremely deformable, but can experience jamming transitions, like cells that form confluent tissues~\cite{park_collective_2016,oswald_jamming_2017}.  Two additional categories of jamming transitions therefore involve packings of \textit{non-spherical} particles and packings of \textit{deformable} particles.

Experiments~\cite{lu_ordering_2016} and numerical simulations~\cite{torquato_jammed_2010-1} of jamming of non-spherical particles, such as ellipsoids~\cite{williams_random_2003}, spherocylinders~\cite{wouterse_effect_2007}, and polyhedra~\cite{chen_complexity_2014}, have probed the validity of the Maxwell criterion for determining mechanical stability~\cite{zeravcic_excitations_2009}.  Generalized to non-spherical particles, the criterion states that static packings need to possess $z = 2 \times DOF$ contacts per particle to be mechanically stable, where $DOF$ is the number of degrees of freedom per particle~\cite{maxwell_calculation_1864}. As discussed earlier, the Maxwell criterion holds for sphere packings, where the number of contacts per particle is exactly $2d$ and the number of degrees of freedom per particle is equal to the spatial dimension $d$. Does the Maxwell criterion hold for non-spherical particles?  Consider a packing of spheroids, i.e., an ellipsoid of revolution with one symmetry axis.  Two degrees of freedom are required to specify the orientation of a spheroid, and three degrees of freedom are required to specify the position of the center of mass.  According to the Maxwell criterion, each spheroid should possess $N_c = 2\times (3 + 2) = 10$ contacts.  However, experiments on spheroid packings clearly show that the number of contacts at jamming onset is always below ten, violating the Maxwell criterion.  Therefore, jammed packings of spheroids are \textit{hypostatic}, and possess fewer contacts than the apparent number of degrees of freedom.  Subsequent numerical and experimental studies have shown that nearly all jammed packings of non-spherical particles are hypostatic.  

Numerical studies have focused on investigating the mechanisms that give rise to hypostaticity in jammed packings of non-spherical particles.  These studies reveal that the number of missing contacts is the same as the number of \textit{quartic modes}~\cite{donev_underconstrained_2007}.  Perturbations along these quartic modes give rise to a quartic increase in the potential energy versus the amplitude of the perturbation in the zero-pressure limit~\cite{papanikolaou_isostaticity_2013-1,schreck_constraints_2012-1,vanderwerf_hypostatic_2018,brito_universality_2018-1}.  Quartic modes are in fact responsible for stabilizing jammed packings of a wide range of non-spherical particles.
\begin{figure}
    \centering
    \includegraphics[width=\columnwidth]{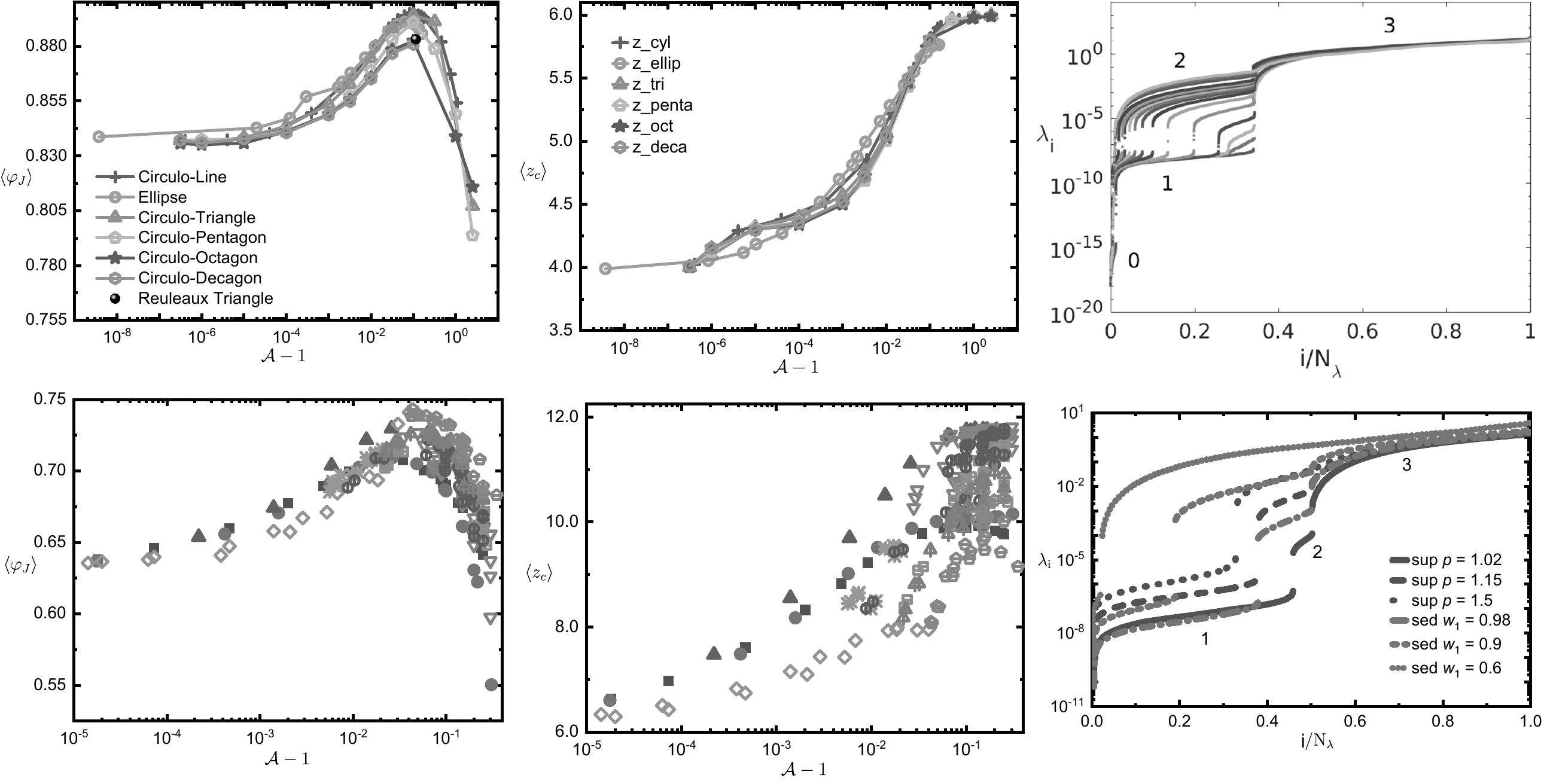}
    \caption{From left to right: Average packing fraction $\langle \varphi_J \rangle$ and excess contact number $\langle z_c \rangle$ at jamming onset as a function of asphericity ${\cal A}-1$ and low-frequency eigenvalues $\lambda=\omega^2$ of the dynamical matrix for jammed packings of non-spherical particles in two (top) and three (bottom) dimensions. The eigenvalues are divided in three branches: (1) quartic modes, (2) rotational modes, and (3) translational modes. Figures reproduced with permission from Refs.~\cite{vanderwerf_hypostatic_2018} \copyright (2018) American Physical Society (top panel) and Ref.~\cite{yuan_jammed_2019} \copyright (2019) Royal Society of Chemistry (bottom panel).}
    \label{fig:asphericity}
\end{figure}

The structural properties of jammed packings of non-spherical particles are summarized in Fig.~\ref{fig:asphericity}.  Both the packing fraction and coordination number for jammed packings of non-spherical particles show non-trivial dependence on the shape parameter ${\cal A}$.  In 2D, $\mathcal{A} = p^2 / 4\pi a$, where $p$ is the perimeter and $a$ is the area of the particles. In 3D, $\mathcal{A} = (4\pi)^{1/3} (3V)^{2/3} / S$, where $V$ and $S$ are the volume and surface area of the particles, respectively. In particular, the average packing fraction and coordination number at jamming onset follow master curves as a function of $\mathcal{A} -1$.  These results suggest that the shape parameter controls the jamming behavior of packings of non-spherical particles. 

\begin{figure}
    \centering
    \includegraphics[width=0.95\columnwidth]{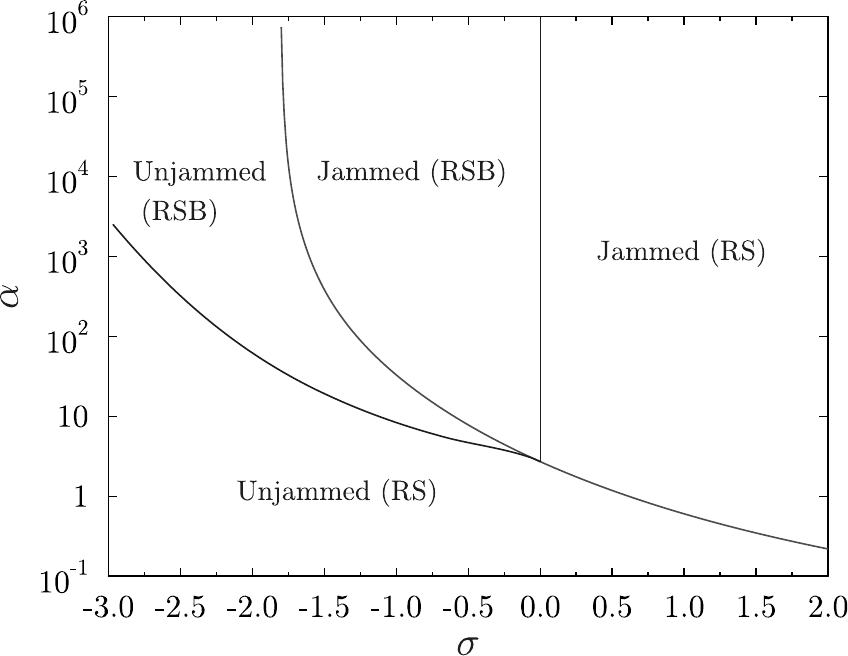}
    \caption{The jamming behavior in the $\alpha$-$\sigma$ plane, which quantify the density and convexity of the obstacles on the $N$-dimensional hypersphere in the generalized perceptron model. Figure reproduced with permission from Ref.~\cite{brito_universality_2018-1} \copyright (2018) National Academy of Sciences of the United States of America.}
    \label{fig:perceptron}
\end{figure}

Using an extension of the perceptron model, Brito {\it et al.}~\cite{brito_universality_2018-1} predicted the scaling of the coordination number versus the pressure and the existence of quartic modes in the VDOS for jammed packings of non-spherical particles~\cite{schreck_constraints_2012-1,vanderwerf_hypostatic_2018}.  The generalized perceptron model predicts the zero-temperature phase diagram for jammed packings of non-spherical particles in the $\alpha$-$\sigma$ plane, where $\alpha$ and $\sigma$ represent the density and convexity of the obstacles on the hypersphere.  (See Fig.~\ref{fig:perceptron}.)  The critical behavior in the presence of spherical asymmetry is substantially altered from its counterpart derived for jamming of spherical particles.  In particular, different scaling exponents are predicted for the gap and force distributions, and the per-particle coordination number scales as
\begin{equation}
\Delta z \sim c_{\mathcal{A}} p \;\;\; \mbox{with} \;\;\; c_{\mathcal{A}} \sim \mathcal{A}^{1/2} \; ,
\end{equation}
and consequently the shear modulus follows the scaling $G \sim p /\mathcal{A}^{1/2}$.  These results were corroborated by previous numerical studies of jammed packings of non-spherical particles~\cite{schreck_comparison_2010,vanderwerf_hypostatic_2018}, proving the versatility of the mean-field theory for jamming. 

Particle shape has also been explored as a tool to incorporate frictional forces into the jamming framework, a direction that requires further experimental investigation~\cite{singh_shear_2020}.  Prior work has investigated jammed packings of particles with circular (in 2D) or spherical (in 3D) asperities on the particle surfaces~\cite{papanikolaou_isostaticity_2013-1}.  These particles have an effective static friction coefficient $\mu_{\rm eff}$ and the packing fraction $\phi_J(\mu_{\rm eff})$ and contact number $z_J$ of packings at jamming onset mimic the behavior found for jammed packings of frictional spherical particles using the Cundall-Strack model~\cite{cundall_discrete_1979}.  However, these packings are isostatic if all contacts between asperities are counted.  Other work ~\cite{ikeda_jamming_2020-2} has also modeled friction by considering rough particle surfaces.  In their studies, the amplitude and frequency of the bumps on the disk surface are continuously varied.  Although this work shows a clear reduction of the coordination number and packing fraction at jamming onset with increasing static friction coefficient, it is limited to small deviations from smooth spherical shapes and it is not clear how to count contacts at concave surfaces.

The nature of the jamming transition in static packings of soft and deformable particles is much less understood.  Experimental studies of aqueous colloidal suspensions have shown that particle deformability plays a crucial role in determining the mechanical properties of colloidal suspensions, leading to glassy behavior with increasing concentration~\cite{mattsson_soft_2009}.  Colloids with larger deformations resemble strong glass-forming liquids.  In contrast, harder colloids display super-Arrhenius increases in viscosity with increasing concentration, typical of fragile glasses.  Other studies have shown that the shape parameter of deformable particles plays an important role in determining the onset of rigidity in cell monolayers~\cite{bi_motility-driven_2016}. 

Including particle deformability in models of jamming can describe a range of biological phenomena, such as wound healing, development of biological tissues, and macromolecular crowding.  The deformable particle model (DPM) studied in 2D~\cite{boromand_jamming_2018} and 3D~\cite{wang_structural_2021}, and the elastic polymer ring model~\cite{gnan_microscopic_2019} in 2D have been developed to describe such systems.  In 3D, the shape-energy function includes a volume term that represents particle compressibility, a surface tension term, and a bending energy term that assigns an energy cost to surface deformations.  Studies of jammed packings of deformable particles using the DPM have shown that deformable particles possess polyhedral shapes with ${\cal A} \sim 1.16$ at confluence, i.e., particles deform until they completely fill all of the available space as observed in epithelia and endothelia~\cite{mongera_fluid--solid_2018,ilina_cellcell_2020}.  Several other models have also studied the onset of jamming in the context of confluent tissues, such as the vertex~\cite{bi_energy_2014} and self-propelled Voronoi~\cite{bi_motility-driven_2016} models.  Here, unjamming transitions are strongly influenced by cell activity, which includes cell speed and persistence of the direction of motion. In recent work by Agoritsas~\cite{agoritsas_mean-field_2021-1}, activity has been incorporated into a Dynamical Mean-Field Theory (DMFT) for a system of infinitely-persistent particles at jamming onset. DMFT represents an important avenue of research for advancing the theory of jamming to driven and active granular systems and it has been recently applied to the perceptron model~\cite{manacorda_gradient_2022}.  While still in its infancy, this direction of research could potentially result in the integration of particle shape degrees of freedom into the mean-field theory of jamming.  
\\
\\
Our goal with this chapter was to summarize the predictions of the mean-field theory of jamming that have been confirmed by numerical simulations and experiments. Most of the confirmations have been for mean-field predictions concerning jamming of frictionless, spherical particles, while more recent work has shown confirmations of the mean-field theory of jamming for frictionless, non-spherical particles and frictional, nearly spherical particles.  We also present current efforts in expanding the mean-field theory to systems that more closely resemble externally driven granular media, cell aggregates, and active colloidal suspensions.  The physics of jamming is far more diverse and rich than the specific topics related to confirmations of the mean-field theory that we presented here.  We direct the reader to more detailed review articles for an exhaustive description of numerical and experimental studies of jamming~\cite{liu_jamming_2010-1,van_hecke_jamming_2010-1,behringer_physics_2019} and to the book by Parisi, Urbani and Zamponi for a complete description of the jamming replica theory~\cite{parisi_theory_2020}.

\section{Acknowledgements}
We would like to thank all of our collaborators in the community without which this work would not have been possible.  Particular gratitude goes to our editor Patrick Charbonneau and Elisabeth Agoritsas, Ada Altieri, Bulbul Chakraborty, Cameron Dennis, Andrea Liu, Peter Morse, Sid Nagel, Nidhi Pashine, Mark Shattuck, Pierfrancesco Urbani, Eric Weeks, and Francesco Zamponi. E.C. and F.A. acknowledge funding from the Simons Collaboration on Cracking the Glass Problem via Award No. 454939 and C.S.O. acknowledges funding from NSF Grant No. DMREF-2118988. 

\bibliographystyle{ws-book-har}

\bibliography{jamming}

\end{document}